\documentclass[fleqn,usenatbib]{mnras}

\usepackage{amsmath}
\usepackage{graphicx}
\usepackage{color}
\usepackage{ulem}
\usepackage{amssymb}
\usepackage{bm}
\usepackage{hyperref}
\usepackage[dvipsnames]{xcolor}

\usepackage{epsf}
\def\Msun{{\rm M_{\odot}}}
\def\msun{{\rm M_{\odot}}}

\def\Mdoted{{\dot M_{\rm Edd}}}
\def\Ledd{{L_{\rm Edd}}}

\newcommand{\ergs}{\ensuremath{\,\mathrm{erg}\,\mathrm{s}^{-1}}}
\newcommand{\gms}{\ensuremath{\,\mathrm{g}\,\mathrm{s}^{-1}}}

\title[ULXs are beamed] {Ultraluminous X--ray sources are beamed}

\author[Jean--Pierre Lasota \& Andrew King]
{Jean--Pierre Lasota$^{1, 2}$ \& Andrew King$^{3, 4, 5}$\\
$^{1}$ Institut d'Astrophysique de Paris, CNRS et Sorbonne Universit\'e, UMR 7095, 98bis Bd Arago, 75014 Paris, France\\  
$^{2}$ Nicolaus Copernicus Astronomical Center, Polish Academy of Sciences, ul. Bartycka 18, 00-716 Warsaw, Poland\\     
$^{3}$ Astrophysics Group, School of Physics \& Astronomy, University of Leicester, Leicester LE1 7RH, UK\\
$^{4}$ Astronomical Institute Anton Pannekoek, University of Amsterdam, Science Park 904, 1098 XH Amsterdam, Netherlands\\
$^{5}$ Leiden Observatory, Leiden University, Niels Bohrweg 2, NL-2333 CA Leiden, Netherlands\\
}

\begin{document}

\date{\today}

\maketitle

\begin{abstract}
We show that magnetar models for ultraluminous X--ray sources (ULXs) have serious internal inconsistencies. The magnetic fields required to increase the limiting luminosity for radiation pressure above the observed (assumed isotropic) luminosities are completely incompatible with the spin--up rates observed for pulsing ULXs. We note that at least one normal Be--star + neutron star system, with a standard (non--magnetar) field, is observed to become a ULX during a large outburst and return to its previous Be--star binary state afterwards. We note further that
recent polarimetric observations of the well--studied binary Cyg X--3 reveal that it produces strong emission directed away from the observer, in line with theoretical predictions of its total accretion luminosity from evolutionary arguments.
We conclude that the most likely explanation for ULX behaviour involves radiation beaming by accretion disc winds. A large fraction of X--ray binaries must pass through a ULX state in the course of their evolution.
\end{abstract}

\begin{keywords}
accretion, accretion discs – black hole physics – binaries: close – pulsars: general – X-rays: binaries.
\end{keywords}

\section{Introduction}

Ultraluminous X-ray sources (ULXs) are defined by the two conditions

(i) apparent luminosities (assumed isotropic) $L_X > 10^{39}\ergs$, and

(ii) locations away from galaxy centres. 

These restrictions select a group of objects not straightforwardly explained either as accreting stellar--mass binaries, or as more massive accretors. Condition (i) requires $L_X$ to exceed the Eddington luminosity for a $10\msun$ black hole, i.e.
%The defining luminosity has no
%astrophysical basis but was chosen for corresponding roughly to the Eddington luminosity of a $10\msun$ black hole, this critical luminosity
%being defined as
\begin{equation}
\label{eq:Ledd}
\Ledd = 1.3 \times 10^{38} m\, {\rm erg\,s^{-1}},   
\end{equation}
%for a $10\msun$ black hole,
with $m={M}/{\Msun} = 10$, which implies a corresponding Eddington accretion rate 
%\begin{equation}
\begin{align}
\Mdoted \equiv \frac{\Ledd}{\eta c^2} & = 1.4\times 10^{18}\eta_{0.1}^{-1} m\,{\rm g\,s^{-1}} \\
& = 2.2 \times 10^{-8}\, \eta_{0.1}^{-1} m\,\Msun\rm yr^{-1},
\end{align}
%\end{equation}
where $\eta=0.1\eta_{0.1}$ is the radiative efficiency of accretion. 
Condition (ii) rules out the central massive black holes in galaxies.

ULXs were identified as a separate class of objects at the end of the previous millenium \citep{Colbert9907}. By now, only two models of ULX behaviour remain under serious consideration.

The older of these two current models for ULX behaviour is disc--wind beaming \citep{King0501}. This asserts that the assumption of isotropic emission made in computing $L_X$ from observations is not valid for
%In the beaming picture ULXs emit most of their flux in narrow beams along the accretion disc axis of 
binary systems transferring mass at rates $\dot m\Mdoted$, with $\dot m \gg 1$, because in this case radiation pressure expels most of the transferred mass in quasispherical winds which are opaque except along narrow channels along the accretion disc axis \citep{Shakura73}. This means that most of the emitted accretion luminosity $\sim \Ledd$ is beamed along these channels.
ULXs are sources where the observer lies in one of the beams: the effect
is that the apparent (assumed isotropic) luminosity inferred is
\begin{equation}
L_{\rm sph} \sim\frac{1}{b}\Ledd \gg \Ledd, 
\end{equation}
where the total solid angle of the two channnels is $4\pi b$.

The more recent model for ULX behaviour, which we shall refer to as the `magnetar model', was inspired by
the discovery by \citet{Bachetti1410} that the source 
ULX--2 in the galaxy M82 is pulsed. This implies that the accretor is a magnetized neutron star\footnote{This discovery had the effect of making an early model for ULXs invoking accretion on to more massive (`intermediate--mass', or `IMBH') black holes relatively unattractive.}.
The magnetar model asserts that unusually strong surface fields of neutron--star accretors   
reduce the electron scattering opacity defining the Eddington luminosity, making $\Ledd$ numerically larger: the high luminosities of ULXs are actually {\it sub}--critical in this picture.
For magnetar--strength fields ($\gtrsim 10^{14}$~G) the modified $\Ledd$ exceeds the assumed--isotropic luminosity $L_X$.

We note that the two models imply fundamentally different significances for the ULX phenomenon.  
Disc--wind beaming asserts that the ULX state is one that a large fraction of otherwise standard X--ray binaries pass through during a particular phase of their evolution, whereas the magnetar hypothesis reduces the ULX class to a relatively small subset of these systems defined by very strong magnetic fields.

The aim of this paper is to evaluate recent evidence allowing a clear decision between 
beaming and strong magnetic fields as the basic cause of ULX behaviour.

\section{Beaming}
\label{sec:beaming}

The suggestion by King et al. (2001) of beaming as the explanation for the high apparent luminosities of ULXs was motivated by study of the X--ray binary Cyg X--2 \citep{King9910}. The neutron star in this system has evidently survived the companion star attempting to transfer $\sim 3\msun$ to it at a highly super--Eddington rate, without retaining more than a small fraction of it. 

This corresponds closely to the picture of how a disc deals with a super--Eddington mass rate suggested by \citet{Shakura73}. A radiation--pressure powered wind from the disc
surface keeps the disc accretion rate at the local Eddington limit corresponding to each disc radius. This raises the true total emitted accretion luminosity only by a logarithmic factor, to  
\begin{equation}
L_{\rm acc} \simeq \Ledd[1 + \ln\dot m],
\label{eq:supereddlum}
\end{equation}
so that even a huge (by X--ray binary standards) accretion rate of $\sim 10^4\Mdoted$ would give a total accretion luminosity of only $10\Ledd$, i.e $\sim 2\times 10^{39}\ergs$ for a $1.4 \Msun$ neutron star.

But importantly the emission is now highly anisotropic:
the outflowing wind is densest near the radius at which the full Eddington luminosity is attained, and has a large optical depth both along the disc plane and in the vertical direction. Thus most of the disc radiation emitted within the wind region diffuses by scattering until it is able to escape through the central open funnels parallel to the disc axis. 
Since the funnel is tall and thin and has scattering walls, the escaping radiation is beamed by a factor $b \ll 1$, so that the apparent luminosity deduced by an observer in the beam, who assumes the luminosity to be isotropic, is
\begin{equation}
\label{eq:superedlbeam}
L_{\rm app} = \frac{1}{b}\Ledd[ 1 + \ln\dot m] \gg \Ledd.  
\end{equation}

\citet{King0902} showed that for $\dot m \gg1$, the observed correlation  $L_{bb} \propto T_{bb}^{-4}$ between ULX soft X--ray blackbody luminosity and temperature
implies that
\begin{equation}
\label{eq:bfactor}
b \simeq \frac{73}{\dot m^2}.
\end{equation}
This agrees with deductions from simple accretion disc theory, as conditions far from the disc centre are set by the mass supply rate, while those near the disc centre all converge to what is set by a near--Eddington central accretion rate.

\citet{King0520} noted that when the accretor is
a magnetized neutron star its magnetic axis is not necessarily aligned with the disc (i.e. funnel) axis, and it is very common for the neutron star spin to be misaligned from the binary orbit defining the accretion disc plane.
When these three axes are not aligned the system appears as a pulsing ULX, or PULX. 
For a neutron--star spin axis strongly misaligned from the central disc axis at the spherization radius, large polar caps produce the sinusoidal pulse light curves observed in PULXs since
a significant part of the pulsed emission can escape without scattering, giving a large pulse fraction.
Using this disc--wind--beaming  model \citet{King1702,King0519,King0520} (see also \citealt{king0623}) were able to obtain self--consistent sets of parameters for the 10 known PULXs, finding magnetic fields in the range of $\sim 2 \times 10^{10} - 10^{13}$G, mass--transfer rates $\dot m$ between $\sim 10$ and $\sim 100$, and beaming factors from $\sim 0.01$ to $\sim 0.5$.

\section{Magnetar models}

But as we noted above, soon after the discovery of the first PULX a different explanation of the apparent super--Eddington luminosities observed in these X--ray sources became possible. 
\citet{Dalosso0515,Eksi0315})
assumed that the PULX magnetic fields had magnetar ($\gtrsim 10^{14}$ G) fieldstrengths. These substantially reduce the scattering cross-sections and so increase the critical luminosity at which the radiation pressure force equals the pull of gravity. In this scenario PULX luminosities are above the usual Eddington luminosity, but actually sub--critical, so that accretion proceeds in the same way as in other X--ray pulsars.

Indeed, very strong magnetic fields lower the Thomson and Compton scattering opacity \citep{Canuto0571,Herold0579} for photons with energies $E_\gamma$ lower than the cyclotron frequency $E_{\rm cyc}$:
\begin{align}
\label{eq:sigmaB}
\frac{\sigma_{B1}}{\sigma_T}\approx &  \sin^2\theta +\left(\frac{E_\gamma}{E_{\rm cyc}}\right)^2\cos^2 \theta\\
\frac{\sigma_{B2}}{\sigma_T}\approx &  \left(\frac{E_\gamma}{E_{\rm cyc}}\right)^2, \, \rm for \, \frac{E_\gamma}{E_{\rm cyc}}\ll 1,
\end{align}
where indices 1 and 2 correspond to the two linear photon polarizations, $\sigma_T$ is the Thomson cross-section and $\theta$ is the angle between the directions of the magnetic field and light propagation. The opacities depend on the photon polarization, but as shown by \citet{Paczynski0792} their Rosseland means differ at most by a factor 2, depending on the angle between
the direction of the photon propagation and the field lines. Therefore in the presence of a very strong magnetic field,
the critical luminosity corresponding to the equality of the radiation pressure and gravitational forces can be written as
\begin{equation}
\label{eq:Lcrit}
{L_{\rm crit}} \approx 2 B_{12}^{4/3}\left(\frac{g}{2\times 10^{14}\rm cm\,s^{-2}}\right)^{-1/3}{\Ledd},
\end{equation}
where $g=GM/R^2$ \citep{Paczynski0792}.
Thus in this picture the apparent (assumed isotropic) PULX luminosities $\gtrsim 10^{40}\ergs$ must be emitted by a plasma permeated by magnetar--strength fields
$> 10^{14}\rm G$. 

Although at first sight attractive, the idea of magnetars in PULXs faces the difficulty that these very strongly magnetized neutron stars have never been observed in binary systems (see \citealt{King0519, king0623} and references therein). Accepting it 
requires belief in a cosmic conspiracy making them detectable in binaries 
only when these have high mass transfer rates. We shall see in the next Section that this idea disagrees with observations in any case.

We now know that out of the $\sim$ 1800 observed ULXs (see \citealt{king0623} and references therein) at least 10 contain magnetized neutron stars, detected through their periodic pulses (PULXs). Four of them are transient: they are members of Be--X binary systems, which become X--ray sources when the eccentric orbit of the compact companion (in most if not all cases a neutron star) of the massive Be star crosses its circumstellar disc. In most cases this disc--crossing produces sub--Eddington--luminosity outbursts (called ``Type I''), but from time to time, most probably because of von Zeipel-Kozai-Lidov oscillations of the circumstellar disc \citep{Martin0914}, it results in a giant (super--Eddington; `Type II') outburst. 

Swift/XRT observations of galaxies NGC 4945, NGC 7793 and M81 suggest that although persistent ULXs dominate the high end of galaxy luminosity functions, the
number of systems emitting ULX luminosities are probably dominated by transient sources. These transients are most probably not Be--X systems \citep{Brightman0723}.

\section{Magnetic fields in ULXs cannot have magnetar strengths}
\label{sec:ULX}

There is a simple physical argument that rules out the presence of magnetars in observed PULXs. The argument is based on the value of their spin-up rate $\dot \nu$ ($\nu$ is the pulsar's spin frequency).

After the discovery of the first PULX M82 ULX-2, \citet{Kluzniak1503} pointed out that it differs from other X-ray pulsars (XRPs) not only through its higher luminosity but also in its extremely high 
spin--up rate. 

\begin{figure}
\begin{center}
\includegraphics[width=\columnwidth]{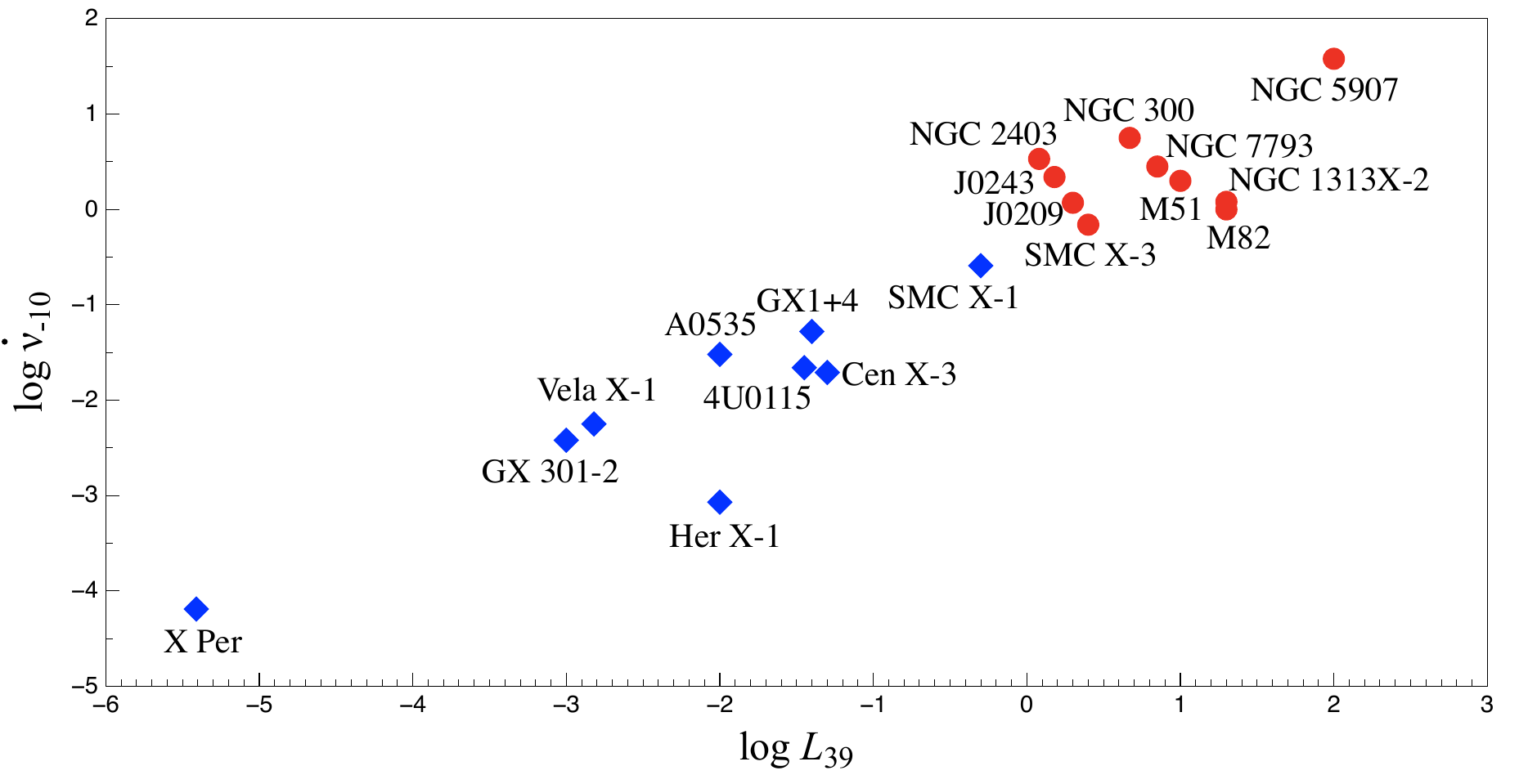} 
\caption{The $L_{39}$ -- $\dot \nu_{-10}$ diagram for XRPs and PULXs. Red dots: the ten PULXs with known spin-up rates. Blue diamonds:
selected (for comparison) sub--Eddington--luminosity X-ray pulsars (For details see \citealt{king0623})}.
\label{fig:dotnul}
\end{center}
\end{figure}

It is immediately obvious that both `normal' 
X--ray  systems and PULXs lie on  
exactly the same strong correlation between 
spin--up rate $\dot \nu$
and X--ray luminosity $L_X$ -- see Fig~\ref{fig:dotnul}. 
This correlation extends more than seven orders of magnitude in luminosity, and arises because the spin--up results from the accretion torque on the neutron star:
\begin{equation}
\dot\nu = \frac{\dot J(R_M)}{2\pi I} = \frac{\dot M (GMR_M)^{1/2}}{2\pi I} \propto \dot M^{6/7}\mu^{2/7},
\label{eq:dotnudef}
\end{equation}
where $R_{\rm M} \propto \dot M^{-2/7}\mu^{4/7}$ (from Eq.~\ref{eq:rm}) is the magnetospheric radius, $\mu = BR^3$ the neutron star's magnetic moment (with $B$ the field and $R$ the neutron--star radius) and $I$ is the neutron star's moment of inertia.

The magnetospheric radius is defined by the equation \citep{FKR2002}
\begin{equation}
R_{\rm M} = 2.6 \times 10^8 q\, \left(\frac{\dot M}{10^{17}\gms}\right)^{-2/7} \left(\frac{M}{\Msun}\right)^{-3/7} \mu_{30}^{4/7}\, \rm cm,
\label{eq:rm}
\end{equation}
where the factor $q\sim 1$ takes into account the geometry of the accretion flow at the magnetosphere and $\mu=10^{30} \mu_{30}\rm Gcm^3$. 

Assuming $M\approx 1\Msun$, $q\approx 1$, $I=10^{45}\rm g\,cm^2$ and using Eq. (\ref{eq:rm}), Eq.~(\ref{eq:dotnudef}) 
gives
\begin{equation}
\label{eq:mdotnudot}
    \dot M \approx 5.7 \times 10^{18} \dot\nu^{7/6}_{-10} \mu_{30}^{-1/3} \gms
\end{equation}
as the accretion rate required to spin up a magnetised neutron star at the rate $\dot \nu$.

Now we can calculate the luminosity produced by this accretion rate. Supercritical luminosities are not proportional to the accretion rate (see Eq.~\ref{eq:supereddlum}).
But  very strong magnetic fields make the critical luminosity  much larger than the Eddington value, i.e.  $L_{\rm crit} \gg \Ledd$ (see Eq. \ref{eq:Lcrit}). So for  $L_{\rm crit} > L_X \gtrsim \Ledd$ the standard formula $L_X = 0.1 \dot M c^2$ applies, even though $L_X$ exceeds the usual Eddington value.

Then for magnetar PULXs ($\mu \gtrsim 10^{31} {\rm Gcm^3}$) we get from Eq. (\ref{eq:mdotnudot}) the luminosity
\begin{equation}
\label{eq:Lxnu}
 L_X \approx 2 \times 10^{38} \dot\nu^{7/6}_{-10} \mu_{31}^{-1/3} \ergs \approx \Ledd.
\end{equation}
But in deriving this equation we assumed $L \gtrsim L_{\rm crit}\gg\Ledd$, which would require a much smaller field (i.e. $\mu_{31} \ll 1$).\footnote{Eq. (\ref{eq:Lxnu}) explains why sub--Eddington--luminosity XRPs ($\mu_{31} \sim 0.001 - 1$) have $\dot \nu < 10^{-10}\rm s^{-2}$.}
This contradiction
shows that magnetars cannot be present in
systems with both $L_X > 10^{39}\ergs$ and $\dot\nu \gtrsim 10^{-10}\rm s^{-2}$.

In other words, PULXs cannot contain magnetars. 

This in turn means that the super--Eddington luminosity
observed in PULXs is not intrinsic, and must presumably be anisotropic, i.e. beamed.

Importantly, since $L_{\rm crit} \sim B^{4/3}$, in a dipole field the critical luminosity decreases radially outwards as $R^{-4}$. So at radius $\sim 100$ stellar radii all of the cross--section suppression is lost, well inside the magnetosphere. 
Then a 
hyper--Eddington luminosity emitted near the 
neutron--star surface would blow away all of the gas in the upper part of the accretion column, thus cutting off the mass supply supposedly producing the posited hyper--Eddington emission. This rules out the interpretion of the CRSF observed in the magnetized, non--pulsing ULX-8 in M51 as an effect of protons orbiting a $9\times 10^{14}$G magnetic field, as envisaged by \citet{Brightman1804}, and provides another strong argument against the presence of magnetars in ULXs\footnote{We are grateful to the anonymous referee of the present paper for suggesting this line of argument}.

\section{Magnetic Fields in PULXs}

We conclude from the last Section that neutron stars in PULXs have magnetic fields spanning the same range as the usual XRPs -- from  $10^8$ G to
several $10^{13}$~G \citep{Revnivtsev18}. They are evidently normal XRPs observed in a special phase of the evolution of their parent binary systems, as is implicit in the original suggestion by 
\citet{King0501}. We can see examples of this in real time
in observations of Be--star PULXs. These are normal XRPs for most
of their lifetimes, and become PULXs only during their occasional giant outbursts. This allows one to follow the transformation of an XRB into a PULX and its return to  `normal' again.
\begin{figure}
\centering
\includegraphics[width=8.0cm,height=6.0cm,angle=0]{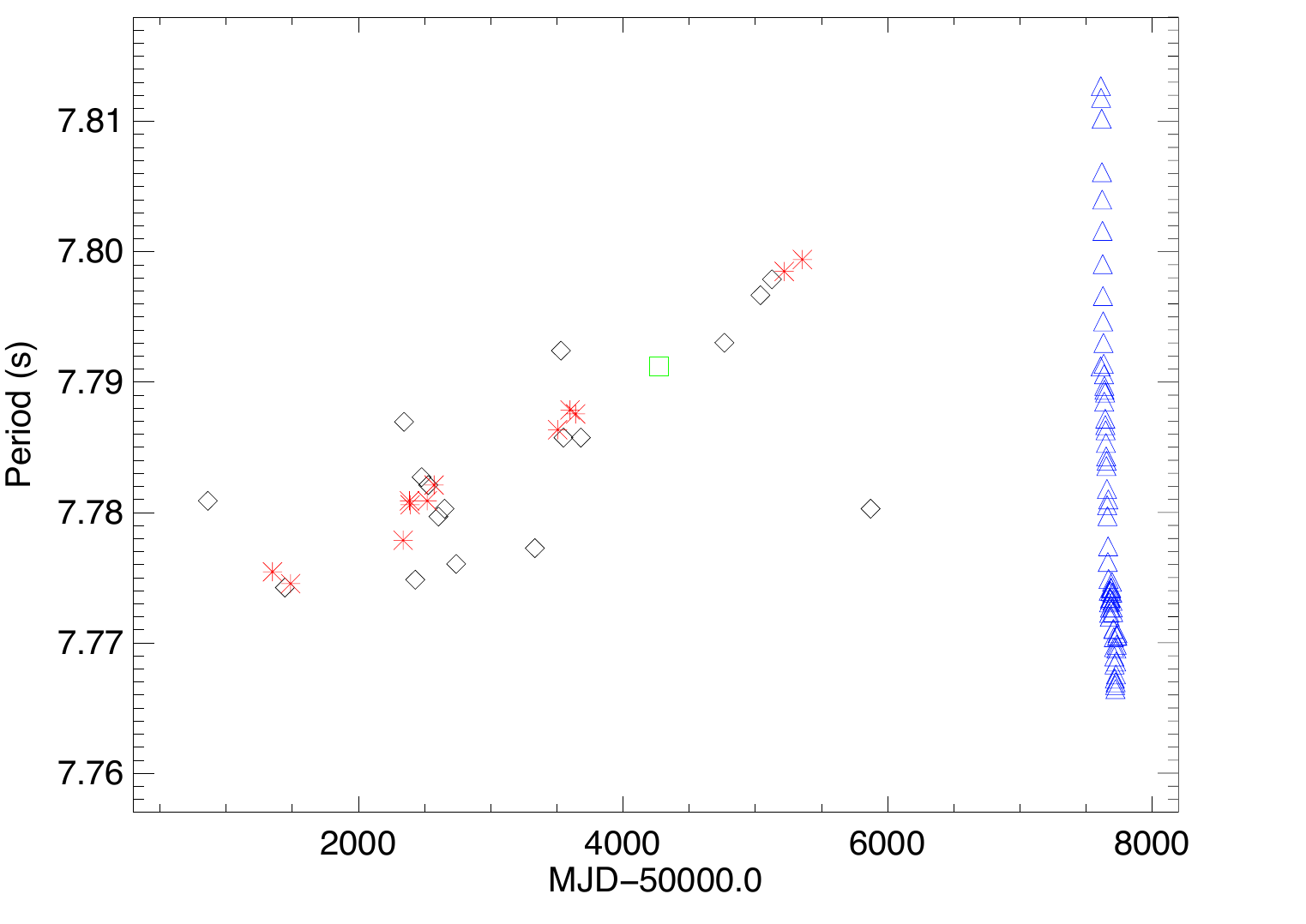}
\caption{X-ray derived pulsed period history of SMC X-3. Black diamonds
and red stars denote RXTE period detections above the 99 and 99.99 per cent
confidence levels respectively. Blue triangles denote Swift detections of the
pulse period during the current outburst. A single XMM–Newton detection
at MJD 54274 was found in the literature and is denoted by a green square. {(From \citealt{Townsend1117}}.)}
\label{fig:smcx3}
\end{figure}

The best studied case is that of the binary SMC X-3. It shows that as the system enters the ULX phase, the neutron--star spin evolution becomes dominated by
the accretion torque, as assumed in Eq. (\ref{eq:dotnudef}). Between giant outbursts this sources is an XRP which spins down.
In Fig.~\ref{fig:smcx3} \citep{Townsend1117} this corresponds to the time right up to the beginning of a giant outburst, on MJD 57599; then a significant spin--up is observed. From SMC X-3's long--term spin history \citet{Townsend1117} deduce that the angular momentum transferred by accretion during the 5--month giant outburst was larger than the total angular momentum lost by magnetic braking over the previous 18 years of the spin--down phase. The long--term 
spin--down rate of SMC~X-3 is about 500 times lower than the rate of spin--{up} observed during the giant outburst, showing that the torques acting during this outburst are far larger than during the out--of--outburst phases. During weaker (Type I) outbursts, the spin period continues to increase, but during the giant outburst the 
spin--up rate is tightly correlated with the X--ray luminosity through the super--Eddington phase \citep{Weng0717}, in agreement with Eq. (\ref{eq:dotnudef}). This means that in PULXs the spin--up rate is strongly correlated with the X--ray luminosity both in time and over the population. There are no magnetars in PULXs.

\section{Cyg X-3: the second hidden ULX in the Galaxy}

Quite recently, \citet{Veledina0323} performed X--ray polarimetry indicating `unambiguously' that the Wolf--Rayet X-ray binary Cyg X--3, consisting of a helium star transferring mass to a black hole on its thermal timescale, is a ULX with a beaming factor\footnote{We use here the symbol $b$ as defined in \citet{King0902}; by contrast \citealt{Veledina0323} use $b$ to denote his $1/b$.} $b\approx 1/65$, but seen from the side. This system is assumed to contain a black hole. Earlier inferred examples of `sideways' ULXs notably include the extreme source SS433
(cf \citealt{Begelman0706,King0116}).

From Eq. (\ref{eq:bfactor}) we find that this requires an Eddington factor $\dot m \simeq 69$. This is consistent with the 
estimates of the mass transfer rate found by \citet{Lommen1105} on evolutionary grounds.
Together with the similar estimates for SS433 (\citealt{Begelman0706,King0116}) this appears to be explicit confirmation that compact binaries with mass transfer rates exceeding the Eddington rate produce beamed emission, as first suggested by \citet{King0501}. Moreover, the estimate (\ref{eq:bfactor}) appears to be in reasonable agreement with observation.

The good match between the luminosities of some ULX nebulae and the luminosity of their ULX irradiators has been used as an argument against
strong geometrical beaming in these ultraluminous sources since in these cases the nebula would see an isotropic emission. However, the sources in question
are spectrally soft and  it is not
surprising that the irradiating luminosity inferred from photoionisation
modelling is consistent with the observed luminosity (see \citealt{king0623} for a detailed discussion).

\section{Conclusion}

We have shown that magnetar models for ULX behaviour have serious internal inconsistencies. In particular the fieldstrengths required to increase the radiation pressure luminosity limit above the observed (assumed isotropic) luminosities are completely incompatible with the spinup rates observed for PULXs. In addition we note that at least one normal Be--star system, with a standard (non--magnetar) field, is observed to become a ULX during a large outburst.

In contrast, recent polarimetric observations of the well--studied binary Cyg X--3 reveal that it produces strong emission beamed away from the observer.

We conclude that ULXs are beamed.

\section{Data Availability}
No new data were generated or analysed in support of this research.

%%%%%%%%%%%%%%%%%%%% REFERENCES 

\bibliographystyle{mnras}
\bibliography{BULX} % if your bibtex file is called example.bib


\end{document}